%% file: main.tex
\documentclass[conference]{IEEEtran}
\IEEEoverridecommandlockouts

\usepackage{cite}
\usepackage{amsmath,amssymb,amsfonts}
\usepackage{graphicx}
\usepackage{textcomp}
\usepackage{xcolor}
\usepackage{tikz}
\usepackage[caption=false,font=footnotesize]{subfig}
\definecolor{NeonRefGreen}{HTML}{39FF14}
\definecolor{RepoLinkBlue}{HTML}{0057FF}
\usepackage[colorlinks=true,linkcolor=black,citecolor=NeonRefGreen,urlcolor=RepoLinkBlue]{hyperref}
\usepackage{cleveref}
\usepackage{orcidlink}
\usepackage{eso-pic}
\usetikzlibrary{arrows.meta,calc,positioning}

\def\BibTeX{{\rm B\kern-.05em{\sc i\kern-.025em b}\kern-.08em
    T\kern-.1667em\lower.7ex\hbox{E}\kern-.125emX}}

\newif\ifanonymouspaper
\ifdefined\anonymousversion
  \anonymouspapertrue
\else
  \anonymouspaperfalse
\fi
\ifanonymouspaper
  \newcommand{\paperPdfAuthor}{Anonymous Authors}
\else
  \newcommand{\paperPdfAuthor}{Markus Baumann, Claudia Linnhoff-Popien, and Jonas Stein}
\fi

\ifanonymouspaper\else
\AddToShipoutPictureFG*{%
  \AtPageLowerLeft{%
    \hspace{\dimexpr(\paperwidth-\textwidth)/2\relax}%
    \raisebox{0.35in}[0pt][0pt]{%
      \parbox[b]{\textwidth}{\scriptsize
        \copyright~2026 IEEE. Personal use of this material is permitted.
        Permission from IEEE must be obtained for all other uses, in any current
        or future media, including reprinting/republishing this material for
        advertising or promotional purposes, creating new collective works, for
        resale or redistribution to servers or lists, or reuse of any copyrighted
        component of this work in other works.}%
    }%
  }%
}
\fi
\hypersetup{
  pdftitle={Symmetry Alone Is Not an Ansatz: Task-Aligned Interactions in Equivariant Quantum Circuits},
  pdfauthor={\paperPdfAuthor},
  pdfsubject={Variational quantum learning and equivariant ansatz design},
  pdfkeywords={quantum machine learning, geometric quantum machine learning, equivariant quantum neural networks, variational quantum circuits, ansatz design, inductive bias, generalization}
}

\renewcommand{\figurename}{Fig.}
\renewcommand{\tablename}{Tab.}
\crefname{figure}{Fig.}{Figs.}
\Crefname{figure}{Fig.}{Figs.}
\DeclareRobustCommand{\IEEEauthorrefmark}[1]{\smash{\textsuperscript{\footnotesize #1}}}

\begin{document}

\title{Symmetry Alone Is Not an Ansatz: Task-Aligned Interactions in Equivariant Quantum Circuits}

\ifanonymouspaper
\author{%
\IEEEauthorblockN{Anonymous Authors}
}
\else
\author{%
\IEEEauthorblockN{%
Markus Baumann\IEEEauthorrefmark{1}\orcidlink{0009-0007-3575-1006}\thanks{Corresponding author: \href{mailto:markus.baumann@campus.lmu.de}{markus.baumann@campus.lmu.de}.},
Claudia Linnhoff-Popien\IEEEauthorrefmark{1}\orcidlink{0000-0001-6284-9286},
and Jonas Stein\IEEEauthorrefmark{1}\orcidlink{0000-0001-5727-9151}
}
\IEEEauthorblockA{\IEEEauthorrefmark{1}\textit{QAR-Lab, Department of Computer Science, LMU Munich, Munich, Germany}}
}
\fi

\maketitle

\begin{abstract}
\input{txt/0_Abstract}
\end{abstract}


\section{Introduction}
\label{sec:introduction}
\input{txt/1_Introduction}

\begin{figure*}[!t]
\centering
\includegraphics[width=.94\textwidth]{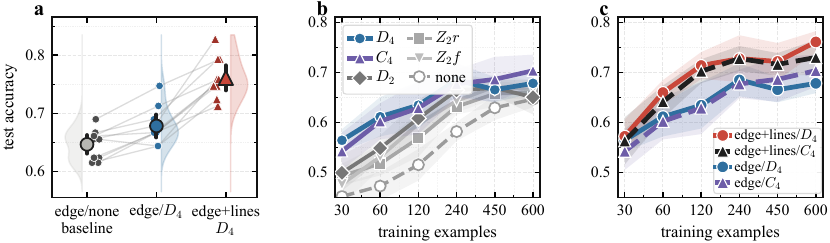}
\caption{Main empirical evidence. (a) Full $D_4$ equivariance improves the original edge ansatz over no sharing. (b) The subgroup sweep across training sizes shows that $C_4$ tracks $D_4$ closely. (c) Adding winning-line interactions to the equivariant edge ansatz improves test accuracy while preserving the imposed symmetry.}
\label{fig:main}
\end{figure*}

\section{Protocol}
\label{sec:protocol}
\input{txt/2_Protocol}

\section{Results}
\label{sec:results}
\input{txt/3_Results}

\section{Discussion}
\label{sec:discussion}
\input{txt/4_Discussion}

\section*{Appendix}
\input{txt/5_Reproducibility}

\ifanonymouspaper\else
\section*{Acknowledgment}
This work was supported by the LMU Sustainability Fund (EfOiE), the German
Federal Ministry of Research, Technology and Space (BMFTR) through QuCUN,
QuaRDS, and CAQAO, the Munich Quantum Valley consortia K5 and K7, and the
Bavarian Ministry of Economic Affairs project 6GQT.
\fi

\bibliographystyle{IEEEtran}
\bibliography{references}

\end{document}

%% file: txt/0_Abstract.tex
The success of variational quantum learning models crucially depends on choosing parametrizations that reflect the structure of the problem at hand. Symmetries provide one of the clearest such structures: whenever transformations of the input leave the desired outcome unchanged, this invariance should be built into the model rather than discovered during training. However, imposing a symmetry does not by itself determine a useful ansatz. Even within the symmetry-preserving space, one must decide where the trainable degrees of freedom should be placed. In this work, we study this remaining design freedom in equivariant variational quantum circuits. Building on symmetry-based parameter sharing, we disentangle two architectural choices: how much symmetry should be enforced, and which symmetry-respecting interactions should be trainable. Using Tic-Tac-Toe as a fully enumerable and structurally transparent test case, we find that suitable subgroups preserve most of the generalization benefit. By contrast, the dominant gains arise from gates acting directly on decisive task motifs. Thus, symmetry defines the admissible design space, while effective ans\"atze require an additional task-informed choice of trainable interactions.

%% file: txt/1_Introduction.tex
A central challenge in variational quantum machine learning is not merely to
build expressive quantum circuits, but to build the right expressive circuits.
Layered parameterized circuits with data re-uploading
~\cite{benedetti2019parameterized,perezsalinas2020data} can be highly flexible,
but flexibility alone is not a design principle. Without architectural
guidance, a model may spend parameters and data on distinctions irrelevant to
the task, while failing to allocate capacity to the structures that determine
the label. The practical question is therefore not only how much expressivity a
quantum model has, but where this expressivity is placed.

Symmetry is one of the few sources of prior knowledge that can be stated
exactly. Consider a target map \(y:\mathcal X \to \mathcal Y\) and a group
\(S\) acting on the input space through transformations
\(V_s:\mathcal X \to \mathcal X\). The task is invariant under \(S\) if
\[
    y(V_s[x]) = y(x)
    \qquad \text{for all } x \in \mathcal X,\; s \in S .
\]
In words, a symmetry specifies transformations of the input that must not
change the correct prediction. A translated image still depicts the same
object, a permuted graph still represents the same graph, and a rotated
Tic-Tac-Toe board still has the same winner. A model that respects this
structure by construction need not rediscover it from data.

This observation is central to geometric deep learning
~\cite{bronstein2017geometric}: classical
architectures such as convolutional neural networks organize trainable degrees
of freedom around task symmetries~\cite{cohen2016group}. In quantum machine
learning, geometric quantum machine learning represents known transformations
on the Hilbert space and constrains circuits accordingly
~\cite{larocca2022group,nguyen2024theory}. Related work has developed
group-invariant models, equivariant quantum neural networks,
permutation-equivariant architectures, graph-equivariant circuits, and
symmetry-informed variational ans\"atze
~\cite{larocca2022group,nguyen2024theory,skolik2023equivariant,sauvage2024building}.

\begin{figure*}[t]
\centering
\includegraphics[width=0.98\textwidth]{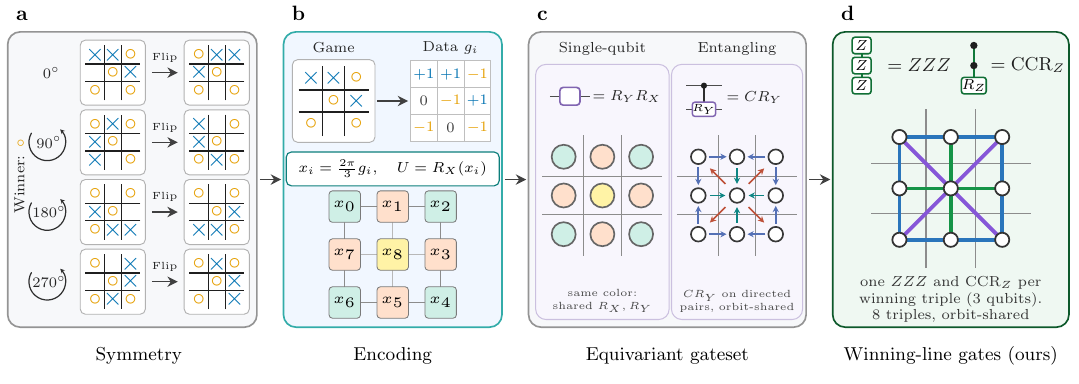}
\caption{From task symmetry to a task-aligned equivariant circuit.
(a) Tic-Tac-Toe labels are invariant under the $D_4$ rotations and reflections of the board.
(b) Board entries $g_i\in\{-1,0,+1\}$ are encoded as $R_X(x_i)$ rotations with $x_i=\frac{2\pi}{3}g_i$. The chosen indexing exposes the corner, side, and center orbits.
(c) The Meyer et al.~\cite{meyer2023symmetry} edge ansatz uses orbit-shared $R_YR_X$ rotations and directed $CR_Y$ entanglers on neighboring and center-connected fields.
(d) Our extension adds orbit-shared $ZZZ$ and symmetric $\mathrm{CCR}_Z$ gates on the eight winning triples, directly matching the label-relevant motifs while preserving equivariance.
Panel layout adapted from Meyer et al.~\cite{meyer2023symmetry}.}
\label{fig:protocol}
\end{figure*}

Meyer et al.~\cite{meyer2023symmetry} gave a concrete blueprint for this
program. If a data embedding \(U(x)\) satisfies
\[
    U(V_s[x]) = U_s U(x) U_s^\dagger ,
\]
then the data symmetry is represented by a unitary action \(U_s\). Trainable
blocks that commute with this representation, together with compatible state
preparation and readout, yield a model whose prediction is invariant by
architecture rather than by optimization.

However, symmetry alone does not specify an ansatz. It tells us which inputs
should be identified, which parameters may be shared, and which observables
should be aggregated, but not where trainable interactions should live. Two
circuits can respect the same group action and still match the task very
differently. We therefore treat symmetry not as a switch, but as a constrained
design space in which one must still choose the sharing strength and the motifs
on which trainable quantum interactions are placed.

Tic-Tac-Toe is well suited for this question. The task is finite,
exactly enumerable, and structurally explicit: the relevant board
symmetries, labels, and decisive three-cell motifs are all known in advance.
This lets us isolate ansatz-design choices rather than infer them from a
large, opaque benchmark.

We separate two choices that are often entangled. The first is how much of the
available board symmetry should be enforced: full equivariance is natural, but
smaller subgroups may preserve most of the useful bias while leaving the
circuit more flexible. The second is which symmetry-respecting interactions
should be made trainable. The baseline captures the board's local geometry,
but the target function is decided by complete winning triples. Our
extension therefore asks whether placing trainable interactions directly on
those motifs improves the equivariant ansatz.

Our results support this design-oriented view. Enforcing symmetry improves the
edge baseline, and suitable subgroup sharing retains most of that benefit. The
largest gain, however, comes from enriching the equivariant circuit with
task-aligned interactions. Thus, symmetry is not merely a constraint that makes
a model smaller. It is a language for organizing circuit design, but the
effectiveness of the model still depends on placing trainable degrees of
freedom where the structure of the task actually lives.

%% file: txt/2_Protocol.tex
Each legal, reachable Tic-Tac-Toe board is represented by
nine values $g_i\in\{-1,0,+1\}$ for circle, empty, and cross.
The three labels are circle win, draw, and cross win, encoded
as $(+1,-1,-1)$, $(-1,+1,-1)$, and $(-1,-1,+1)$. As in
Meyer et al.~\cite{meyer2023symmetry}, unfinished games are
counted as draws. Since the full set contains only 5478 legal,
reachable boards, we enumerate it exactly. The labels
are invariant under the eight rotations and reflections of the
board, forming the dihedral group $D_4$, as illustrated in
Fig.~\ref{fig:protocol}(a). Train and test splits
are class-balanced and orbit-disjoint: a fixed test set of 348
boards is held out as complete $D_4$ orbits, and training
examples are drawn only from the remaining orbits, so no test
board is related to a training board by a board symmetry.

We use one qubit per field, as shown in Fig.~\ref{fig:protocol}(b).
Each value $g_i$ is mapped to the angle $x_i=2\pi g_i/3$ and
encoded by an $R_X$ rotation,
\[
U(g)=\bigotimes_{i=0}^{8} R_X\!\left(\frac{2\pi}{3}g_i\right).
\]
The fields are numbered cyclically around the rim, with the
center last. This makes the three $D_4$ position orbits explicit:
corners $\{0,2,4,6\}$, sides $\{1,3,5,7\}$, and center $\{8\}$.

We separate the imposed symmetry from the choice of gates.
Let $H\subseteq D_4$ be the subgroup that we enforce. For
$h\in H$, let $V_h$ act on the board and let $P_h$ be the
corresponding qubit permutation. The encoding is equivariant:
transforming the board before encoding is equivalent to
encoding first and then permuting the qubits,
\[
U(V_h g)=P_h U(g) P_h^\dagger .
\]
A trainable block $W_H(\theta)$ preserves this symmetry if
\[
P_h W_H(\theta) P_h^\dagger = W_H(\theta)
\qquad \text{for all } h\in H .
\]
In practice, we enforce this by sharing parameters across gates
whose supports lie in the same $H$-orbit. For directed gates,
the order of the support is kept fixed, so control and target
roles are not accidentally exchanged. Thus the symmetry fixes
the sharing pattern, but not the gate family itself.

The circuit uses repeated data re-uploading layers. Each layer
applies the encoding $U(g)$ followed by $p$ repetitions of a
trainable block. Unless stated otherwise, we use $L=3$ layers
and $p=2$ repetitions, keeping depth fixed across all sweeps.

Our baseline is the edge ansatz of Meyer et al.,
shown in Fig.~\ref{fig:protocol}(c). It applies local $R_YR_X$
rotations and directed controlled-$R_Y$ entanglers on neighboring
rim fields and selected center connections. Symmetry enters only
through parameter sharing: gates in the same $H$-orbit share the
same parameter or parameter tuple.

The readout averages Pauli-$Z$ expectations over the three
position orbits,
\[
r_O(g)=\frac{1}{|O|}\sum_{i\in O}\langle Z_i\rangle_g ,
\]
for corners, sides, and center. These three values are used as
class scores. Since the readout orbits are invariant under all
subgroups considered here, the readout is compatible with every
symmetry sweep.

This gives our first design axis: the amount of symmetry imposed.
We keep encoding, edge wiring, gate family, depth, and readout
fixed, and vary only the subgroup $H$. With $r$ denoting a
$90^\circ$ rotation and $f$ a reflection, we compare
$\{e\}$, $\langle r^2\rangle$, $\langle f\rangle$,
$\langle r\rangle=C_4$, $\langle r^2,f\rangle=D_2$, and
$\langle r,f\rangle=D_4$. For the edge ansatz at $L=3,p=2$,
these variants have $204$, $108$, $126$, $60$, $78$, and $54$
trainable parameters, respectively.

The second design axis is the interaction structure. The edge
ansatz couples neighboring fields, but Tic-Tac-Toe labels are
decided by complete three-cell winning lines. We therefore add
orbit-shared gates on the eight winning triples: three rows,
three columns, and two diagonals, as shown in
Fig.~\ref{fig:protocol}(d). Each triple receives one $ZZZ$ rotation
and one $\mathrm{CCR}_Z$ interaction. The resulting edge+lines
model keeps the original edge ansatz but adds trainable
interactions directly on the label-relevant motifs. At
$L=3,p=2$, the full-$D_4$ edge model has $54$ parameters, while
the full-$D_4$ edge+lines model has $90$.

%% file: txt/3_Results.tex
\begin{figure*}[t]
\centering
\includegraphics[width=.94\textwidth]{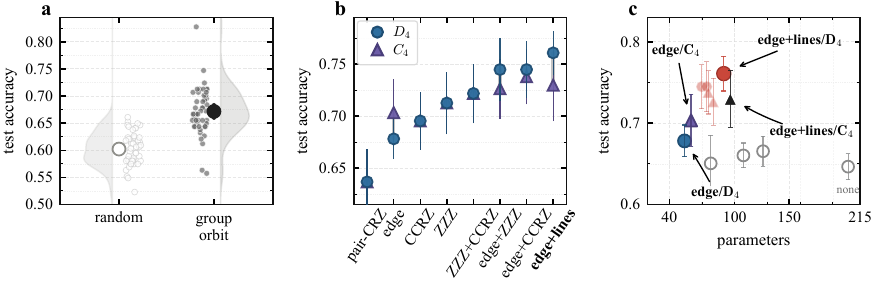}
\caption{Controls and ablations. (a) Parameter-matched random sharing underperforms group-orbit sharing. (b) Ablations are anchored by the edge ansatz. The strongest model combines edge interactions with both line channels. (c) Accuracy is not explained by parameter count alone.}
\label{fig:controls}
\end{figure*}

We first compare no sharing, partial subgroup
sharing, and full $D_4$ sharing with the edge circuit fixed,
then add winning-line interactions and use controls to
separate symmetry-aware sharing from parameter reduction.
All runs use mean-squared error, Adam with learning rate
$0.01$~\cite{kingma2015adam}, batch size 15, 30 minibatch
updates per epoch, 100 epochs, ten random seeds per
configuration (bands and error bars show 95\% confidence
intervals across seeds), PennyLane
simulation~\cite{bergholm2018pennylane}, and PyTorch
optimization~\cite{paszke2019pytorch}.

Across these experiments, three trends emerge, summarized in
Fig.~\ref{fig:main}. First, the unconstrained edge ansatz has more
parameters but generalizes worse than the $D_4$-equivariant
edge model. At 600 training examples, edge/none reaches about
$0.65$ test accuracy, while edge/$D_4$ reaches about $0.68$
with less than half the train-test gap. This supports the basic
equivariance effect: the correct board symmetry improves
generalization.

Second, the symmetry effect is not strictly all-or-nothing.
Full $D_4$ remains the natural symmetry choice, but $C_4$
rotation sharing tracks it closely across the data grid and is
slightly ahead at the largest training sizes. Symmetry strength thus acts as a tunable architectural
bias: partial symmetry preserves most of the benefit while
leaving the circuit less constrained. The random-sharing
control in Fig.~\ref{fig:controls}(a) confirms that the gain is not
caused by parameter reduction alone. Parameter-matched sharing
without group-orbit structure performs worse.

The largest improvement comes from changing the interaction
structure. Adding winning-line gates to the equivariant edge
circuit, as shown in Fig.~\ref{fig:main}(c), raises the best $D_4$
model from about $0.68$ to about $0.76$ test accuracy, with
the $C_4$ variant reaching about $0.73$. The edge ansatz can only
represent rows, columns, and diagonals indirectly, whereas the
line gates act directly on the motifs that determine the label.
This is the central result: symmetry helps, but task-aligned
equivariant interactions help more.

The controls in Fig.~\ref{fig:controls} rule out simpler
explanations. The effect is not just parameter count: edge+lines/$D_4$ has 90 parameters,
far fewer than the 204-parameter unshared edge baseline, yet
performs better. The ablations show that line channels help on
their own, but the strongest model combines edge interactions
with both winning-line channels.

%% file: txt/4_Discussion.tex
We have investigated what remains to be designed once a variational quantum learning model has been constrained to respect a data symmetry. Equivariance is a powerful inductive bias, but it is not a complete ansatz prescription. It tells the model which inputs to treat as equivalent and how to share parameters across symmetry-related parts of the data, but it leaves open how trainable capacity should be assigned to the structures of the task. Our results show that this allocation is a central part of quantum model selection.

Tic-Tac-Toe exposes this point because its symmetries and decisive winning triples are known exactly. Full $D_4$ is the natural symmetry of the board, and enforcing it improves the standard edge ansatz over an unconstrained circuit. At the same time, the subgroup sweep shows that equivariance is not all-or-nothing: in our setting, $C_4$ sharing recovers almost the same generalization behavior as full $D_4$. Thus, the strength of symmetry enforcement can itself be treated as an architectural hyperparameter. Partial symmetry can preserve most of the useful bias while avoiding some of the rigidity introduced by the full group. The choice also affects trainability: symmetry restrictions can speed up training and mitigate barren plateaus~\cite{sauvage2024building,cerezo2023simulability}, yet strongly constrained circuits may approach classically simulable regimes~\cite{cerezo2023simulability}.

The larger effect comes from the choice of interactions. The edge ansatz respects neighboring board fields, but the label is generated by complete rows, columns, and diagonals. Adding orbit-shared $ZZZ$ and $\mathrm{CCR}_Z$ gates on these winning triples keeps the model equivariant while placing trainable interactions directly on the motifs that define the task. The random-sharing, ablation, and parameter-count controls show that the improvement is not explained by fewer parameters or by adding gates indiscriminately. It comes from aligning the equivariant circuit with the geometry of the target function.

Our contribution is therefore a refinement of the symmetry program for variational quantum learning. A practical recipe is to represent the data symmetry, choose the subgroup to enforce, and place trainable interactions on label-relevant motifs. Tic-Tac-Toe is small and exactly enumerable, but precisely for this reason it exposes this principle without confounding factors. Future work should study softer versions of this recipe, such as regularized deviations or horizontal-gate constructions that interpolate between strict equivariance and fully free circuits~\cite{wiersema2025horizontal}, compare winning-line gates with parameter-matched gates on non-winning triples to isolate motif alignment, and apply motif-aligned equivariant ans\"atze to larger tasks where the relevant structures must be inferred from domain knowledge.

%% file: txt/5_Reproducibility.tex
The
\ifanonymouspaper
\href{https://anonymous.4open.science/r/qc_symmetry-5FB4/}{anonymous project repository}
\else
project repository at \url{https://github.com/eybmits/vqml-symmetry}
\fi
contains the code, checked result tables, figure scripts, and manuscript
sources needed to reproduce the paper.